# DESIGN AND FABRICATION OF THE SUSPENDED HIGH-Q SPIRAL INDUCTORS WITH X-BEAMS

M.C. Hsieh*, D.K. Jair[1], and Y.K. Fang[2] and C. S. Lin[3]

*Department of Electronic Engineering, Kun Shan University, Taiwan, R.O.C
[1]Department of Mechanical Engineering, Kun Shan University, Taiwan, R.O.C
[2] Institute of Microelectronic, Department of Electrical Engineering, National Cheng Kung University, Taiwan, R.O.C
[3]Department of Electronic Engineering, Fortune Institute of Technology, Kaoshiung, Taiwan

## ABSTRACT

In this paper, deep sub-micron CMOS process compatible high Q on chip spiral inductors with air gap structure were designed and fabricated. In the design the electromagnetic solver, SONNET, and the finite element program, ANSYS, were used for electrical-characteristics and maximum mechanical strength, respectively. The copper wires were capped with electroless Ni plating to prevent the copper from oxidizing. A $Si_3N_4$/ $SiO_2$ X-beam was designed to increase the mechanical strength of the inductor in air gap. The enhancement of maximum mechanical strength of a spiral inductor with X-beams is more than 4500 times. Among these structures, the measured maximum quality factor (Q) of the suspending inductor and frequency at maximum Q are improved from 5.2 and 1.6GHz of conventional spiral inductor to 7.3 and 2.1 GHz, respectively.

## I. INTRODUCTION

On-chip spiral inductor has been widely used in integrated radio-frequency (RF) CMOS circuits as on-chip matching network, inductive load, passive filter, transformer, etc. [1]. For these applications, the most issue is the large silicon substrate losses [2], which will degrade the quality factor (Q) more. In the past, many high performance structures on Si substrates have been proposed to overcome this issue [1][2][3]. In this work, we propose an alternative technique to solve dilemma. That is using the mature CMOS technology compatible processing and air gap structure to lower the cost in manufacture and reduce substrate losses [4]. However, in preparing of the suspended on-chip inductor with
deep CMOS technology compatible processes, the thickness of inductor's metal wires is thinner than that used with MEMS (microelectric mechanic system) technology. Hence, the structure stability of the suspended inductors, i.e., the inductor's maximum mechanical strength force becomes one of the important concerns, which has been always neglected in the suspended inductors with MEMS technology.

In this work, on-chip spiral inductors were designed and developed to balance the requirements between the electrical characteristics and the structural stability for various applications. Additionally, to tie in requirement of advanced CMOS technology, we used the copper (Cu) to replace aluminum (Al). However, the copper process possesses a serious issue in preparing the air gap spiral inductor due to the oxidation of copper and then damaged in air. Some one has suggested use deposited thin Al layer to cap the copper for relieving this issue [5]. But, in practice, the bottom side of metal wire cannot be deposited for air gap structure. In this letter, using electroless Ni to surround the copper after the air gap etching to prevent copper from oxidizing was developed.

## II. DESIGN AND FABRICATION OF THE SPIRAL INDUCTOR

Although the substrate coupling loss of the suspended on-chip inductor can be reduced by removing embraced $SiO_2$. However, with deep sub-micron CMOS technology, the thickness of metal wire is very thin (0.4um), so that the mechanical stability of the suspended structure, especially for the inductor with large number of turns for some special applications such as cellular phone even after package, may still be vulnerable or distorted by it's own gravity and becomes another serious concern. The evaluations were also done with the software ANSYS 7.1 The simulated maximum impact force as a function of number of turns are plotted in Fig.1. The maximum impact force is defined as the force to cause $1\mu$ m deflection of the suspended metal wire. Since as shown in Fig.2, after having deflected $1\mu$ m, the suspended wires will contact to the lead wire of inductor and induce malfunction. As expected, the maximum impact force is decreased with increasing of number of turns. Moreover, the enhancement in maximum impact force for the X-beam supported inductor is more than 4500 times at 10 turns. The simulation results are useful in design of airgap distance and number of spiral turns. From the result, we can find the maximum impact force for the suspended inductor supported by X-beam (about $1\times10^{-10}$ Nt at 10-turns) is more than $10^4$ times higher than the inductor only supports by pillar (about $2\times10^{-14}$ Nt at 10-turns).

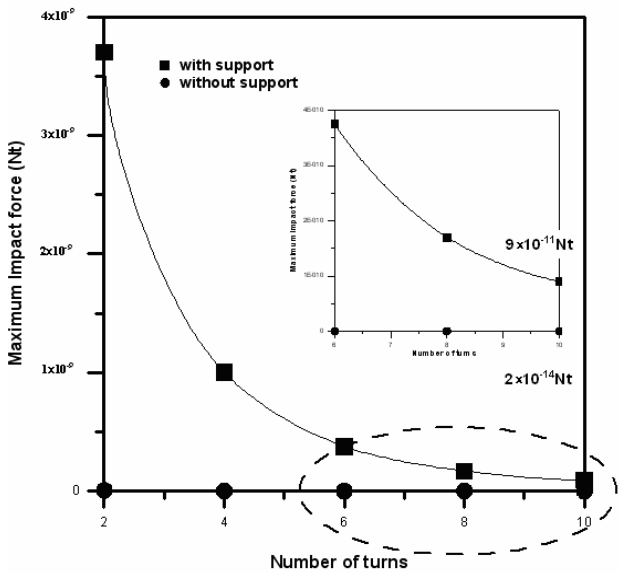

Fig.1 The simulated maximum impact force based on ANSYS 7.1 of the developed inductor for various number of turns. The insert zooms in parts of the simulation results indicated by dot circle.

The samples in this study were prepared with $0.13\,\mu$ m full eight-level Cu metal/FSG (fluorinated oxide) low-k IMD CMOS process technology on silicon substrate with a field oxide of $0.4\,\mu$ m thickness. Fig.2 shows the schematic cross sections of the spiral on chip inductor with air gap and conventional structures. The dimensions of the proposed suspending spiral inductor are $100\,\mu$ m, $10\,\mu$ m, $2\,\mu$ m and 10 turns for the inner diameter, width, spacing, and number of turns, respectively. After having finished the single layer inductor with a thickness of 1um by dual damascene electroless copper plating with a thickness of $0.9\,\mu$ m, the windows of the inductor were opened over the proposed on-chip inductor by removing LPCVD $Si_3N_4$ ($0.1\,\mu$ m) and $SiO_2$ ($0.6\,\mu$ m) layers on the using the process for the opening of bonding pad window. Then the slope wet etching solutions consists of 107ml DI water, 509ml 10:1 BOE, 35ml 49% HF, and 349ml $CH_3COOH$ per one liter at $25^0$C was applied for three minutes to form the air gap, i.e., to remove the $SiO_2$ embraced the inductor as shown in Fig.3.

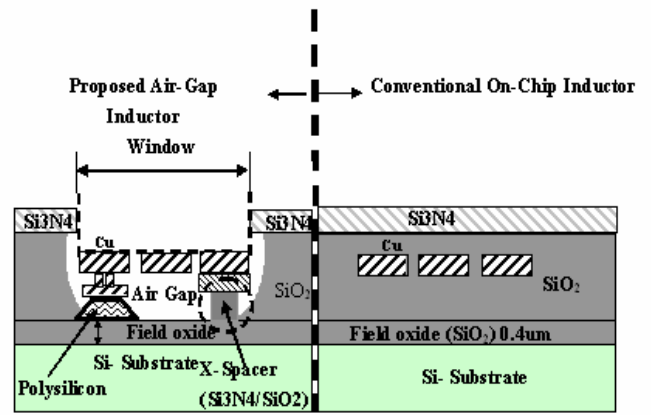

Fig.2 The schematic diagram of cross sections for the on-chip spiral inductor with air gap structure (left) and conventional structure (right)

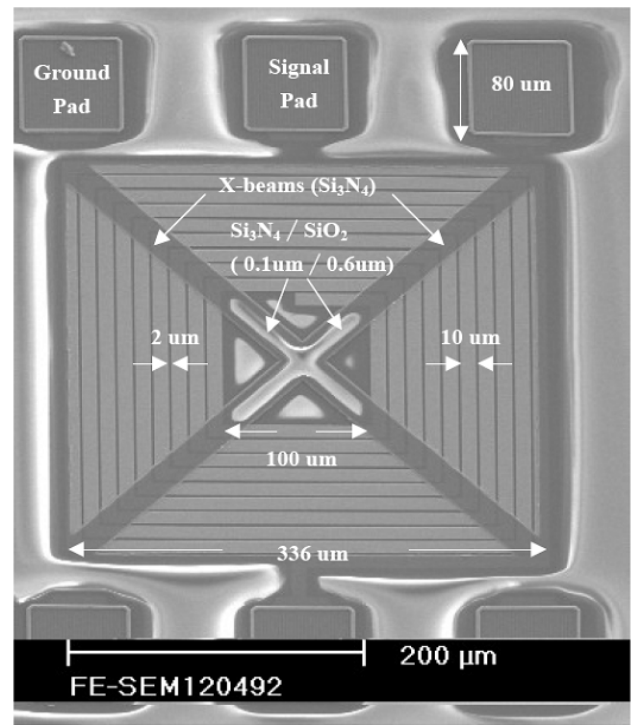

Fig.3 The top view of the completely developed inductor with pads.

As shown in fig.4, the thickness of air gap between metal wire and Si substrate is $2.5\,\mu$ m and $1.6\,\mu$ m between the lead wire and Si substrate. Since the used slope etcher has extremely low etching rate to both Cu, and $Si_3N_4$ [6], the effect of the etcher on the patterns in the other regions, such as metal bonding pads, poly-silicon layer and $Si_3N_4$ passivation layer is very small and can be neglected. Additionally, to enhance the mechanic strength, a crossed spacer beam (called X-beam) was built underneath the Cu wires before the plating of Cu wires as shown in Fig. 3 for detail. The X-beams consist of $Si_3N_4$ and $SiO_2$ layers. Next, some samples were split for Ni electroless plating to prevent the Cu from oxidation and corrosion. Finally, the samples with and without Ni plating were taken to HP 8510C for measurement of S-parameter to obtain Q. The substrate resistivity is 2 K ohm-cm and the measured inductance is 23.3 nH at 1.7 GHz.

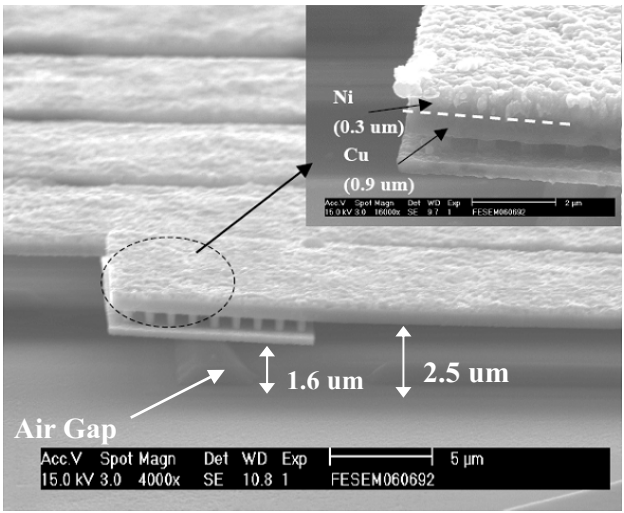

Fig.4 The SEM photo of the suspended spiral inductor without Ni plating and after sloped wet etching.

## III. RESULT AND DISCUSSION

Fig. 3 shows the top view SEM photo of developed inductor with pads before air gap etching. The inductor is a 2-port square spiral. The X-configuration with $Si_3N_4$ (0.1 $\mu$m)/SiO2 (0.6 $\mu$m) in the photo is called X-beams. The sloped wet etched spiral inductor after electroless Ni plating are presented in Figs.4. It shows no CuOx on the top of Cu wire.

Next, in general, Q is defined as,

$$Q = \frac{-\mathrm{Im}\left[\frac{1}{Y_{11}}\right]}{\mathrm{Re}\left[\frac{1}{Y_{11}}\right]} = -\frac{\mathrm{Im}[Y_{11}]}{\mathrm{Re}[Y_{11}]}$$

where Im〔$1/Y_{11}$〕 and Re 〔$1/Y_{11}$〕are the imaginary and real parts of the reciprocal two-port Y-parameter $Y_{11}$, which have been transformed from the de-embedding results of the two-port S-parameters. The major purpose of using the de-embedding is to remove the undesired parasitic effects resulting from ground-signal-ground (GSG) metal pads. Firstly, a dummy pattern should be constructed. The open dummy has the same pattern as the complete pattern for measurement except that the inductor is withdrawn. After transforming S-matrix of measured complete pattern (including both inductor and pads) and open dummy into Y-matrix, respectively, the desired Y-matrix of inductor ($Y_{de\text{-}embed}$) is equal to the Y-matrix of the complete pattern ($Y_{complete}$) minus the Y-matrix of open dummy ($Y_{open}$),

$$Y_{de\text{-}embed} = Y_{complete} - Y_{open}.$$

Experimental results such as characterization with SEM analysis and Q measurement before and after two months after air gap etching on the samples, evident this way is very well for passivation. Fig. 5 shows the measured Q as a function of frequency with and without

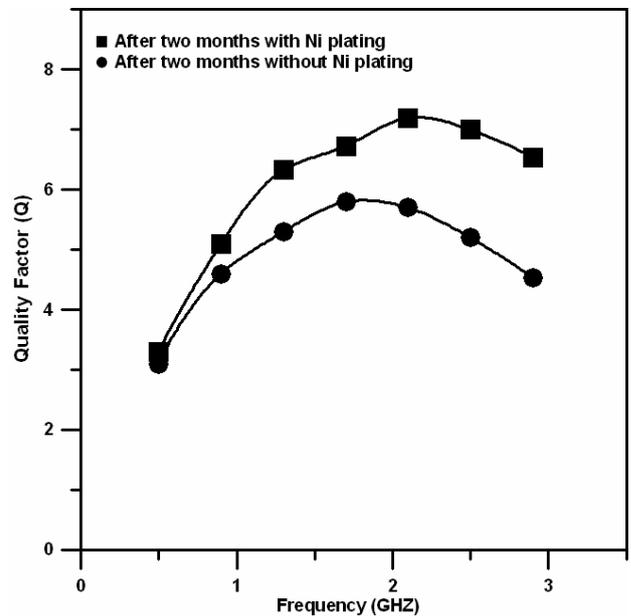

Fig.5 The measured Q of the developed inductor with and without Ni plating after two month in air as a function of frequency. Right side also shows the measured inductances of the developed inductor with Ni plating after two month in air as a function of frequency.

Ni plating after two months in air with 25 $^0$C in temperature and 60% in humidity. The measured maximum Q and frequency for maximum Q just after forming air gap, with and without electroless Ni plating are the same of 7.3 and 2.1 GHz, respectively. Which are almost 40% improvements in the magnitude and 31% in the frequency of the maximum Q in comparison to the conventional $SiO_2$ spiral inductor at 5.2 and 1.6 GHz, respectively. The significant improvement is attributed to the reducing of the capacitances from the metal layer to the substrate and the fringing capacitance between the metal lines. Since the effective dielectric constant of the air is about one fourth (1/4) of the $SiO_2$, thus almost 75% capacitances can be lowered with the air gap dielectric [6]. Clearly, the maximum Q of the inductor with the Ni plated and air gap is almost no degradation at 7.2 for 2.1 GHz in the air. In contrast, without the Ni plating, the maximum Q and frequency at maximum Q are degraded to 5.8 and 1.7 GHz, respectively. This means the Ni plating not only can protect the Cu metal to be oxidized effectively but also don't degrade the maximum Q.

Since the inductors are composed of many strips, the stiffness κ can be investigated with a simple beam. Given a cantilever beam with a length of L, a width of $W_b$, and a thickness of $T_b$, the stiffness κ in the direction $T_b$ is given by $\kappa = \frac{EW_b T_b^3}{4L_b^3}$ [7], where E is the Young's modulus of Al and Cu are 74.14GPa and 130GPa [8,9], respectively. The stiffness κ of the outer beams of suspended spiral inductor is 0.015N/m and the inner is 0.56N/m. Supposed the device undergoes a shock that amounts to an acceleration a= 20g, where g is gravitation acceleration, or 9.8m/s$^2$. Such a force will cause the outer strip of spiral to bend by b=F/κ~0.47μm and 0.002μm. The spacing between the strips is 2μm and the thickness of air gap is 2.5μm, the change in the geometry of the inductor will be negligible. Additionally, the mechanical resonant frequency of the inductor can be roughly calculated by $f_R = \frac{1}{2\pi}\sqrt{\frac{\kappa}{m}}$, we find the $f_R$ of spiral inductor is 3.2kHz. Given the environment vibrations are generally much smaller than 1 kHz, such vibration should not affect the inductor [7].

### 4. SUMMARY

A deep-submicron CMOS process compatible and electroless Ni plated Cu on-chip inductor with air gap and X-beam structure has been successfully developed. Experimental results show that the developed air gap structure and use copper wire promote the inductor's performances very significantly. For example, the maximum quality factor of the suspended inductor is promoted from 5.2 at 1.6 GHz for conventionally inductor to 7.3 at 2.1 GHz. Additionally, the Ni plating nicely prevents the Cu metal from oxidation, and X-beam support improving the mechanical stability more than 4500 times at 10 turns thus enhancing its applications for RF IC profoundly. With the design, experimental results show that the measured maximum quality ($Q_{max}$) and peak-Q frequency of the inductor with air gap were improved to 8.8 and 1.7 GHz, respectively, which present almost 9% improvement in the magnitude and 54% in the peak-Q frequency in comparison to a conventional $SiO_2$ solenoid inductor of 8.1 and 1.1 GHz, respectively.

### Acknowledge

The work was financially supported by the National Science Council under contract NSC 95-2221-E-168-039-